\documentclass[aps,preprint,floatfix,showpacs,superscriptaddress,footnote]{revtex4}
\pdfoutput=1

\usepackage{amsmath}
\usepackage{amssymb}
\usepackage{amsthm}
\usepackage{dcolumn}
\usepackage{epsfig}
\usepackage{graphics}
\usepackage{graphicx}
\usepackage{slashed,epsfig}
\usepackage{longtable}
\usepackage{color}

\definecolor{darkgreen}{rgb}{0,0.5,0}
\definecolor{purple}{rgb}{0.5,0,0.5}
\definecolor{nblue}{rgb}{0.0,0.0,0.50}
\definecolor{scarlet}{rgb}{1.0,0.2,0}
\usepackage[colorlinks=true, pdfstartview=FitV, linkcolor=purple, citecolor= purple, urlcolor=blue]{hyperref}

\newcommand{\beq} {\begin{equation}}
\newcommand{\eeq} {\end{equation}}
\newcommand{\beqa} {\begin{eqnarray}}
\newcommand{\eeqa} {\end{eqnarray}}

\begin{document}

{\par\raggedleft \texttt{SLAC-PUB-15352}\par}
\bigskip{}

\title{Possible multiparticle ridge-like correlations
\\ in very high multiplicity proton-proton collisions}

\author{James~D.~Bjorken} \affiliation{SLAC National Accelerator Laboratory,\\
Stanford University, Stanford, California 94309, USA}
\author{Stanley~J.~Brodsky} \affiliation{SLAC National Accelerator Laboratory,\\
Stanford University, Stanford, California 94309, USA}
\author{Alfred Scharff Goldhaber} \affiliation{SLAC National Accelerator Laboratory,\\
Stanford University, Stanford, California 94309, USA}
\affiliation{C.N. Yang Institute for Theoretical Physics, State University of New York, Stony Brook, NY 11794-3840, USA}

\begin{abstract}
{The CMS collaboration at the LHC has reported a remarkable and unexpected phenomenon in very high-multiplicity high energy  proton-proton collisions:  a positive correlation between two particles produced at similar azimuthal angles, spanning a large range in rapidity.  We suggest that this ``ridge"-like correlation may be a reflection of the rare events generated by the collision of aligned flux tubes connecting the valence quarks in the wave functions of the colliding protons.  The ``spray" of particles resulting from the approximate line source produced in such inelastic collisions then gives rise to events with a strong correlation between particles produced over a large range of both positive and negative rapidity.  We suggest an additional variable that is sensitive to such a line source which is related to a commonly used measure, ellipticity.}
\end{abstract}


\maketitle


\section{Introduction}

An unexpected finding by the CMS collaboration at the Large Hadron Collider (LHC)  is a same-side ``ridge" in two-particle correlations produced in very high-multiplicity proton-proton collisions.  To be precise, plots of the two-body correlation in relative azimuthal angle $\delta\phi$, integrated over relative pseudo-rapidity $\delta\eta$ at a fixed range of transverse momentum $p_\perp$, exhibit a tendency to peak at $\delta\phi \simeq  0$ \cite{Khachatryan:2010gv,Velicanu:2011zz}.

Two immediate qualifications are in order:   First, there is a much higher ridge near $\delta\phi\sim \pi$.  Secondly, for small pseudo-rapidity difference there is a large peak in the correlation function near $\delta\phi=0$.  Both of these effects have plausible explanations.   The away-side ridge reflects the conservation of transverse momentum:   If one particle appears with positive transverse momentum along the $x$ axis, say, then there must be a particle or particles with negative transverse momentum to balance it.  The peak at small $\delta\phi$ for small $\delta\eta$ reflects the presence of a jet and/or a  resonance which decays to more than one particle near a given $\eta$ and $\phi$.

The same-side ridge found by CMS in events with more than 110 particles is statistically significant, but it is not an overwhelming effect.   It comes against a background of gradually increasing correlation with increasing $\delta \phi$.   Thus an alternative interpretation could be that, instead of a ridge, the dynamically important phenomenon is actually a dip at a particular small, but nonzero, value of $\delta\phi$.   A number of authors have made comments and suggested  explanations for the same-side ridge, including \cite{Alderweireldt:2012kt}-\cite{Dusling:2013oia} .

As the authors of the experimental paper have  noted, the ridge recalls a similar, more conspicuous, effect seen in nucleus-nucleus collisions.  A plausible explanation for the nuclear  effect is that when two identical nuclei collide at intermediate impact parameter, the region of overlap between the two has an approximately elliptical shape in the plane transverse to the collision axis.  Simple hydrodynamics then implies that the expansion along the short axis of the overlap ellipse will be more vigorous than the expansion in the long direction, with equal correlations for $\delta\phi\approx 0$ and $\delta \phi \approx \pi$.
This effect is amplified by looking at multiparticle correlations which emphasize the second moment of the azimuthal distribution.

Our purpose here is to suggest that a similar ridge effect, albeit with a quite different physical mechanism, may be at work in the high multiplicity $pp$ collisions.  If we think of a proton as three quarks bound by color forces, then a plausible description of the binding involves color flux `strings'  or tubes connecting to each quark, and to each other, like the bisectors of a triangle with the quarks at the vertices (``Y" diagrams).  The idea of such flux tubes goes back at least to the work of Isgur and Paton \cite{IP}.  Planar networks of strongly interacting gluons connecting valence quarks  are also natural in the higher Fock states of the light-front wave functions of the proton as discussed by Mueller~\cite{Mueller:2002pi}. Such saturating configurations can lead to BFKL phenomena, color-glass models~\cite{McLerran:1993ni}, and they could even provide the color-confining potential of valence quarks as seen in the AdS/QCD soft-wall model~\cite{Erlich:2005qh} using light-front holography~\cite{deTeramond:2005su,Brodsky:2013ar,Brodsky:2006uqa} .   We also note that  $`H'$-diagrams (e.g., two vertical gluons, interacting via horizontal rungs, connecting a heavy quark and antiquark)  which enter the heavy-quark potential~\cite{Appelquist:1977tw} have a similar origin.

\section{Consequences of the collisions of oriented QCD flux tubes}

A linear configuration of the proton has two valence quarks near one end of a gluon flux tube and a third valence quark at the other end. We assume that the flux tube has a transverse size no larger than a few tenths of a Fermi \cite{Buniy:2012yz}.  The probability of finding such a linear configuration is arguably quite considerable, especially given the phenomenological success of quark-diquark models in spectroscopy and other QCD applications. We guess the odds of having such a linear configuration as $20\% $, with an uncertainty either way of a factor two or so. For the collisions of interest, we need the projection of the tube onto the ``impact-plane", and  well over half of the linear configurations should survive this projection.

We are interested in the head-on collision of two such configurations.  The overlap would be significant as long as the impact parameter between the flux tubes were less than the width of either flux tube.  Just as with the elliptical overlap between two nuclei, this shape would tend to produce larger momenta in the transverse direction perpendicular to the lines than in the direction parallel to the lines.
The cost in aligning them azimuthally is a factor 10 or 20, just from simple geometrical considerations. The impact parameter must of course be small, with a transverse size of order of $0.2$  fermi. Along the string direction we can loosen this demand to, say, $0.5$ fermi. This gives a cross-section of about 1 mb. After the penalty factors which we have enumerated are included, the estimated cross-section for the collision of aligned flux tubes  is reduced to a value in the neighborhood of 1 microbarn.  This is quite similar to the cross section for events with a trigger of more than 110 charged-particle tracks used in the CMS experimental `ridge'      analysis \cite{Khachatryan:2010gv}.

\begin{figure}[h]
\centering
\includegraphics[width=14cm]{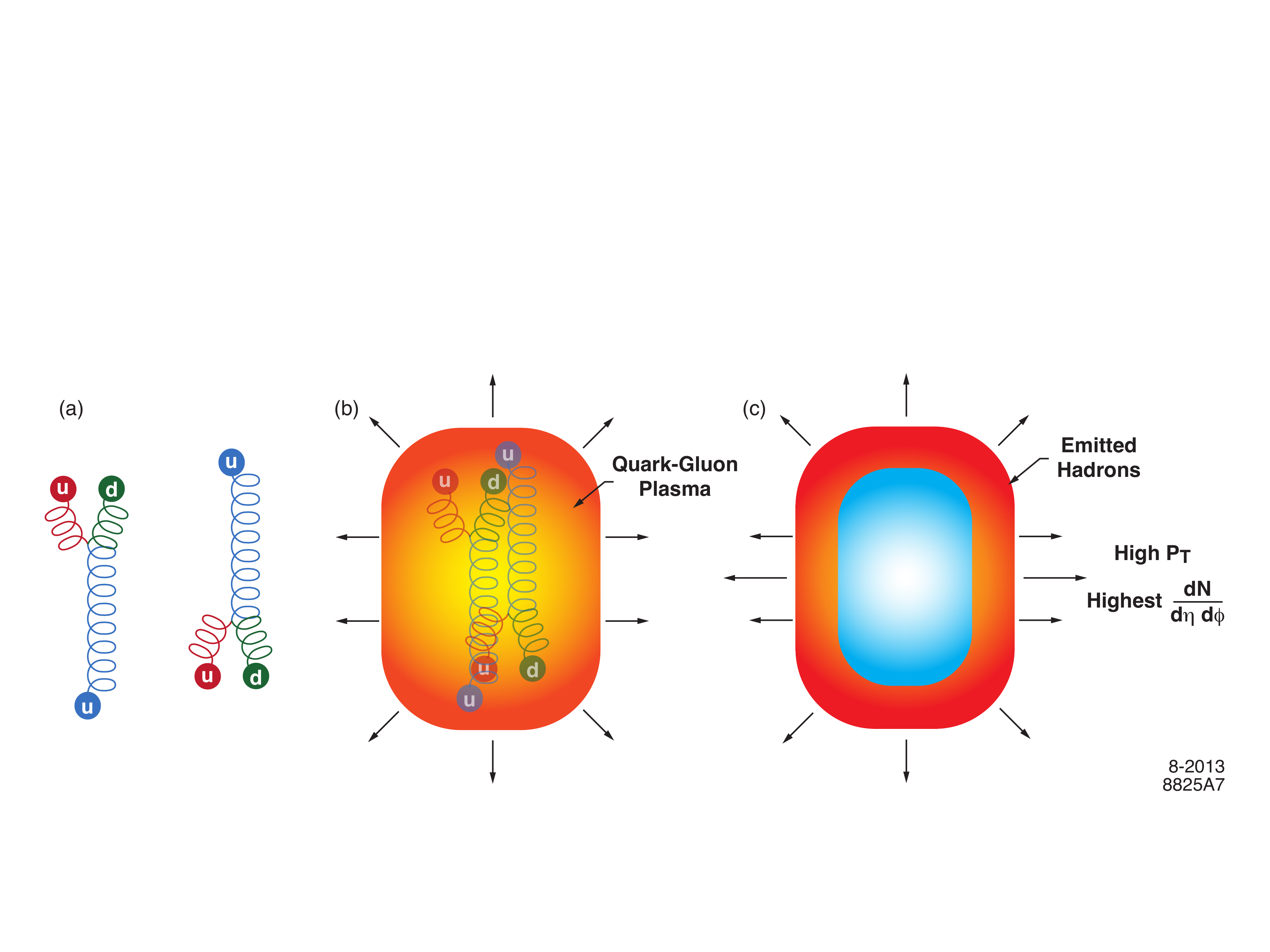}
 \caption{(a) Collisions of oriented  flux tubes   (high-density gluonic strings ) acting between a quark and diquark configuration in  each proton  projectile in $p p$ inelastic collisions.  The tubes are shown projected onto the impact plane, which is perpendicular to the velocities of the two protons.  As drawn, the tubes would miss each other, giving negligible contribution to the ridge signal.
 (b) When the flux tubes overlap, the plasma produced by the collisions of the line sources moves  horizontally, and is subject to two-dimensional hydrodynamic expansion, whereas the hadronic density  moving vertically is subject to the usual three-dimensional expansion.
 (c) Most of the energy density in the transverse direction is emitted as hadrons adjacent to the leading edge of the (transverse) pulse.
 }
\label{Fig1}
\end{figure}

An illustration of the collision of oriented flux tubes -- i.e., high-density gluonic strings -- is given in fig. \ref{Fig1}.

When such a high multiplicity event occurs, almost all of the center-of-mass energy is expected to be expended in particle production; relatively little should be radiated down the beam directions. We estimate roughly  that  $dE_T / d \eta $ in the central region of rapidity could be of order $10-20$ GeV for these events:  The transverse dimensions of the overlapping flux tubes correspond to $1-2$ GeV of momentum.   If one has independent emissions for every piece of the flux tube of length equal to the transverse dimension, then it is reasonable that the resulting total transverse energy per unit rapidity would be an order of magnitude higher.
While some of this transverse energy can be expected to be in visible minijets, it is quite possible that at least some subset of these events are relatively jet-free.  We come to this conclusion by assuming that jets should come from overlap of `occupied pixels' in the colliding flux tubes, where the pixels have dimensions about an order of magnitude smaller than the width of a flux tube.  To obtain the assumed total energy one finds only a minority of the pixels are occupied, and thus the frequency of mini-jet formation may be low.

Looking down the beam direction, the generator of particle production in the flux-tube collisions is an approximate line source, more extreme in azimuthal asymmetry than the almond-shaped source region typical of intermediate-impact-parameter heavy-ion collisions. Therefore, one should expect that this special class of events has even stronger ``ellipticity" effects than in the nuclear case. In particular, a broader transverse momentum distribution is to be expected in the azimuthal direction normal to the flux tubes. Furthermore, this property is expected to be ``boost invariant":  If one were to compare the produced-particle distribution for such an event at zero rapidity at the LHC with what would be seen at zero rapidity for a hypothetical asymmetric collider---say 350 TeV against 35 GeV---the result would be expected to be very similar because the geometry of the relevant active collision area is essentially the same. Therefore the  azimuthal asymmetries seen in the positive-rapidity half of the detector should be strongly correlated with those seen in the negative-rapidity half of the detector.  This feature is the basic correlation structure of all ``ridge" phenomena.

This discussion suggests a flexible search strategy for uncovering not only this phenomenon, but also  related  effects associated with the nontrivial impact-plane geometry.  For example, one might begin by restricting attention to the properties of events in the positive-rapidity half of the detector. After a choice of judicious, creative cuts and/or event scans, one will hopefully have a sample of candidate events exhibiting ``ridge" effects. For each such event, one then looks at the data from the other half of the detector. If the ``ridge" effects we have proposed are genuine, they should be largely reproduced in the other half of the event.  In other words,  the correlations should be seen along virtually the entire rapidity plateau.

In addition to the general strategy described above, we can suggest  a more specific test.
The key point is that a multi-particle correlation should give a much more conspicuous signal than the two-particle correlation used so far in the experimental analysis, but of course only in that small fraction of the events where the prerequisite conditions of coincidence of narrow strings in the projectile and target  are in fact  obtained.   To be specific, we suggest looking at the following vector $\vec V$,  computing its magnitude for each event.  If the number of events with large magnitude are greater than expected from chance, one would have powerful evidence for the proposed colliding flux tube mechanism.
Define
\begin{equation}
{\vec V}=\sum_{i=1}^N [\cos2\phi_i {\hat x}+\sin 2\phi_i{\hat y}] \ \ \ ,
\end{equation}
and obtain the distribution of ${\vec V}^2$.  If the particles were distributed randomly in $\phi$, then the expectation value of ${\vec V}^2$ would be $N$, where $N$ is the number of particles in the event in the given region of transverse momentum.   The probability of getting a value $N^2$ may be estimated by introducing quadrants in the variable $2\phi$:  Assume each vector can take only the values $\pm {\hat x}$ or $\pm {\hat y}$, with each having a probability $1/4$.  Suppose the first vector is $+{\hat x}$.   Then the chance that the remainder would all be in the same direction would be $(1/4)^{N-1}$.  For $N=5$, this would yield a probability $1/256$.   If, among events in which the ridge was seen, with more than 110 particles per event, and 5 particles separated from each other by about one unit in $\delta\eta$ in an interval of $p_\perp$ between 1 and 2~GeV/c, as many as 2\% of the events should show
${\vec V}^2\approx 25$, that could be evidence for the kind of correlation we suggest.   This exercise is equivalent to asking the probability  -- assuming complete randomness in $\phi$ -- that all 5 particles are in either of two opposite octants of $\phi$.  If they were more collimated than that, the probability would be even smaller.

Counting {\underline all }particles in each event in the specified range of transverse momentum, regardless of rapidity separation, should give a reliable measure of the correlation.  Technically, ${\vec V}^2$ is just the square of the usual ellipticity variable.  An advantage of squaring is that maximal ellipticity events are easy to pick out.   Also, it is easier to think about such a scalar variable rather than a vector variable.

At this point let us take a step back to gain perspective on what could cause such phenomena.  Obviously projectile and target must overlap in impact parameter to some extent.   Dynamics, in the form of the conservation of momentum or of the attraction of outgoing particles to each other, can -- as earlier discussed -- produce certain correlations, such as the away-side ridge and the clustering of particles with small relative velocity.   However, the dynamics cannot explain the long-range (in rapidity) near-side correlations.  For this one needs geometry in the transverse plane.  We have given a specific example of such a geometry, the presence of parallel overlapping strings with a substantial projection onto the transverse plane.  Clearly the overlap mentioned for lower-energy nucleus-nucleus collisions also is an example, even though it involves a very different mechanism.

\section{Conclusions}

We have suggested that the ridge-like correlation found by the CMS collaboration~\cite{Khachatryan:2010gv,Velicanu:2011zz}  in high-multiplicity events in $pp$ collisions at the LHC could be evidence for the collision of aligned high-intensity flux tubes connecting the valence quarks of the colliding protons.  A key property is the correlation of ridge events between positive and negative rapidity. Further studies clearly  are warranted,  including an extended analysis of the CMS data based on   the squared correlation vector ${\vec V}^2$.
If one finds evidence for very strongly aligned ``elliptical flow" in $pp$ collisions, where the formation mechanism for the ``ellipse" is in our opinion very different from the geometrical overlap mechanism seen   in nucleus-nucleus collisions, then the  observation of  the ``ridge"  phenomenon in $pp$ collisions  over a large range of rapidity would be a dramatic demonstration of  high density effects in QCD.

\acknowledgements

This research was supported by the Department of Energy  contract DE--AC02--76SF00515.

 \end{document}